\documentclass[%
 reprint,
 amsmath,amssymb,
 aps,
prx,
]{revtex4-2}

\usepackage{graphicx}
\usepackage{dcolumn}
\usepackage{bm}
\usepackage{multirow}
\usepackage{cleveref}
\crefname{equation}{Eq.}{Eqs.} 
 
\usepackage{braket}

\usepackage[table,xcdraw]{xcolor} 

\DeclareMathSymbol{\shortminus}{\mathbin}{AMSa}{"39}
\DeclareMathOperator*{\Motimes}{\raisebox{-0.25ex}{\scalebox{1.2}{$\bigotimes$}}}
\begin{document}

\title{Fast initialization of Bell states with Schrödinger cats in  multi-mode systems }

\author{Miriam Resch$^1$}
\author{Ciprian Padurariu$^1$}
\author{Björn Kubala$^{1,2}$}
\author{Joachim Ankerhold$^{1}$}

\affiliation{$^1$Institute for Complex Quantum Systems and IQST, University of Ulm, 89069 Ulm,
Germany\\
$^2$ German Aerospace Center (DLR), Institute of Quantum Technologies, 89081 Ulm,
Germany
}%

\begin{abstract}
Schrödinger cat states play an important role for applications in continuous variable quantum information technologies. As macroscopic
superpositions they are inherently protected against certain types of
noise making cat qubits a promising candidate for quantum computing. It has been shown recently that cat states occur naturally in driven
Kerr parametric oscillators (KPOs) as degenerate ground states with
even and odd parity that are adiabatically connected to the respective lowest two
Fock states by switching off the drive. To perform operations with
several cat qubits one crucial task is to create entanglement between
them. 
Here, we demonstrate efficient transformations of multi-mode cat states through adiabatic
and diabatic switching between Kerr-type Hamiltonians with degenerate ground state manifolds. These transformations can
be used to directly initialize the cats as entangled Bell states in contrast to initializing them from entangled Fock states.

\end{abstract}

\maketitle

\section{\label{sec:introduction}Introduction}
Superpositions of coherent states commonly referred to as "Schrödinger cat states" are essential in a variety of applications in quantum computation and the simulation of quantum systems.
Recently, they have gained interest as candidates for bosonic encoding of qubits, relying on non-local encoding of the qubits in a superposition of macroscopically distinct classical states to enhance protection against local decoherence \cite{cochrane1999macroscopically,milul2023superconducting}.
Particularly superconducting resonators have become promising platforms for the implementation of such continuous variable cat qubits.  Different realizations have been implemented such as engineered two photon loss \cite{berdou2023one,lescanne2020exponential,leghtas2015confining,reglade2024quantum}, two photon driving \cite{puri2017engineering,grimm2020stabilization,wang2019quantum,iyama2024observation}, and more elaborate schemes combining both approaches \cite{gravina2023critical,gautier2022combined} to confine the qubit subspace and showing long decoherence times \cite{berdou2023one,reglade2024quantum} as well as significant error suppression \cite{lescanne2020exponential,mirrahimi2014dynamically}.

In Hamiltonian-confined cat qubits carried out in Kerr parametric oscillators (KPOs) fast and bias-preserving gates have demonstrated their prospects for quantum computing operations  
\cite{ruiz2023two,puri2020bias,kanao2022quantum,iyama2024observation,hoshi2025entangling,goto2016universal,chono2022two,chono2025high,xu2022engineering,masuda2022fast}. Such Kerr cat qubits are realized by applying a two photon drive to a Kerr resonator and thus creating effectively a double well potential where cat states occur naturally as degenerate eigenstates with defined parity.
Such cat states, which form a protected computational subspace, can be conveniently created by an adiabatic parity-preserving connection to Fock states at zero drive. Moreover, the initialization can be sped up by employing optimal adiabatic control, making the generation of cat qubits highly efficient \cite{xue2022fast}.

One crucial ingredient for successfully performing quantum computing operations between cat qubits is the efficient formation of entanglement between them%
, i.e., between cat states defined in different modes. Previous proposals aim to achieve this either by first creating entangled Fock states which are then adiabatically connected to the respective entangled cat states \cite{hoshi2025entangling}, or by adiabatically initializing unentangled cat states, followed by the application of an entangling gate \cite{hoshi2025entangling}. However, the former approach suffers from exposure to relatively strong decoherence of entanglement in Fock space, whereas the latter approach is limited by the fidelity of the entangling gate. It would thus be a great advantage to start in a pure Fock state and introduce the entanglement-generating interaction during the adiabatic evolution, such that entangled cat qubit states can be initialized adiabatically, fully exploiting the protected computational subspace.

Here, we demonstrate how to directly connect disentangled Fock states to entangled Bell cat states by following a protocol based on a tunable cross-Kerr coupling between two cat qubits, as well as a two-mode mixing drive. All four Bell cat states can be created by adiabatically increasing the single-mode drives and the two-mode mixing drive, while the cross-Kerr interaction is kept constant. Notably, in order to decouple the now-entangled cat qubits, the cross-Kerr interaction can be instantly switched off, without diminishing the fidelity of the Bell cat states. This protocol easily scales up as shown explicitly for $N$ coupled qubits (realized in $N$ KPOs), where it allows us to generate multi-mode cats of the kind $\ket{\alpha_1,...\alpha_N}\pm\ket{\shortminus\alpha_1,...\shortminus\alpha_N}$. 

\section{Creation of cat-qubit Bell states}
\label{sec:cat_qubit}

We start by briefly reviewing the operation of a single KPO as a Kerr-cat qubit \cite{puri2017engineering,grimm2020stabilization}, before introducing the effect of cross-Kerr interaction between two KPOs.

\subsection{Single KPO}
\label{sec:single_KPO}

For a resonator with intrinsic Kerr nonlinearity the Hamiltonian in a frame rotating with the transition frequency $\omega_{01}$ between the first two Fock states is given by $H = -K a^{\dagger 2} a^2$. The corresponding Fock states $\ket{0}$ and $\ket{1}$ form a degenerate eigenstate manifold with eigenvalue zero.
These Fock states can be adiabatically connected to cat states by turning on an external parametric drive resonant with twice the frequency $\omega_{01}$ (two photon drive) with amplitude $\epsilon$. In the same rotating frame, the Hamiltonian of the driven KPO follows as
\begin{align}
    H = -K a^{\dagger 2} a^2 + (\epsilon  a^{\dagger 2} + \epsilon^*  a^2)
    \label{eq:H}
\end{align}
which can be rewritten to
\begin{align}
    H /K= -\left(a^{\dagger 2}-\alpha^{* 2}\right) \left(a^2-\alpha^2 \right)+ |\alpha|^4
\label{eq:H_k}
\end{align}
with $\alpha^2= \epsilon/K$. In the sequel, we use the branch cut implied by the definition $\alpha = \sqrt{r/K}e^{i\phi /2}$, where $\epsilon = r e^{i\phi}$.

Apparently, the manifold spanned by the two coherent states,  $\ket{\pm \alpha}$, forms a degenerate eigenspace with energy $K|\alpha|^4\equiv |\epsilon|^2/K$. This eigenspace is adiabatically connected to the eigenspace formed by the Fock states $\ket{0}$ and $\ket{1}$. Since the Hamiltonian (\ref{eq:H_k}) preserves the photon number parity, it follows that Fock state $\ket{0}$ is adiabatically connected to the cat state $\ket{C^+_\alpha}$, while Fock state $\ket{1}$ is connected to cat state $\ket{C^-_\alpha}$ with the cat states, 
\begin{align}
    \ket{C^\pm_\alpha} = \left(\ket{\alpha}\pm \ket{-\alpha}\right)/N^\pm_\alpha, N^\pm_\alpha= \sqrt{2\left(1\pm e^{-2|\alpha|^2}\right)}
\end{align}
 being eigenstates of the KPO defined in the even and odd Fock state parity manifolds \cite{puri2017engineering}.

The two cat states $\ket{C^\pm_\alpha}$ form the basis of a continuous variable qubit, termed the cat qubit \cite{puri2017engineering, grimm2020stabilization}. It benefits from long lifetimes due to the small phase-space overlap between the two components of the cat states which

can be exponentially suppressed by increasing the amplitude $\alpha$. 
Such cat qubits can be manipulated in two ways, namely, either in the Fock state basis \cite{hoshi2025entangling} at $\alpha=0$, although at the cost of loosing the protection mentioned above, or in the cat space by introducing an additional detuning or an additional driving term and therefore lifting the degeneracy between the ground states, however, at the expense of introducing leakage to higher energy states \cite{puri2017engineering, puri2020bias, grimm2020stabilization}. Single qubit gates can also be realized based on the topological phase that is generated during the rotation of the two photon drive \cite{puri2020bias}.

\subsection{Kerr-coupled KPOs}
\label{sec:coupled_KPOs}

\begin{figure*}	
	
        \includegraphics[width=0.7\linewidth]{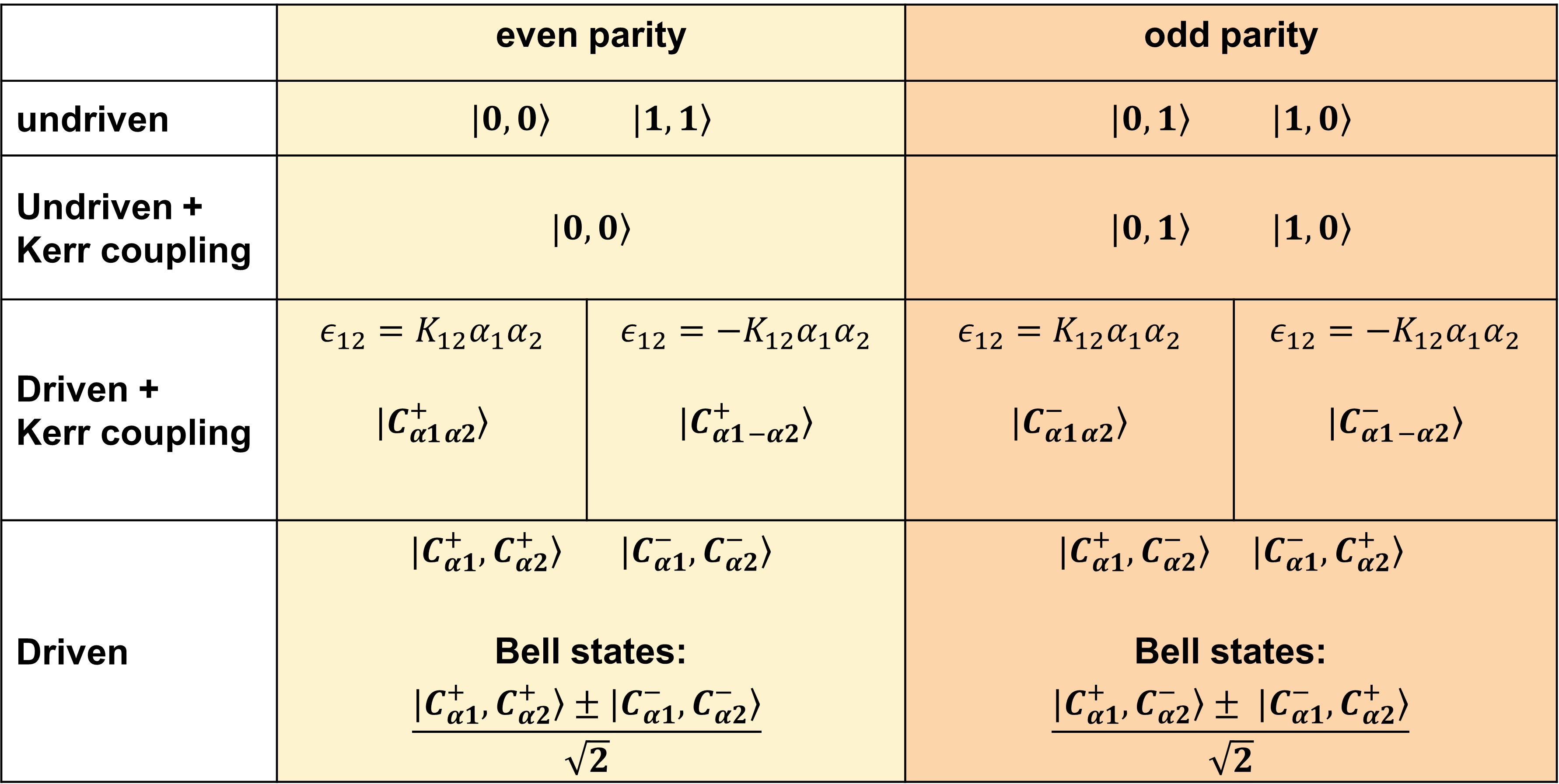}

	\caption{
     Degenerate eigenstates of the two-mode Kerr parametric oscillator depending on the drives and coupling between the modes. The undriven system (top row) is described by the Hamiltonian in Eq.~\eqref{eq:H_full} for $\epsilon_1=\epsilon_2=\epsilon_{12}=0$ and $K_{12}=0$. The driven system (last row) is uncoupled, $K_{12}=\epsilon_{12}=0$.}
\label{fig:eigenstates_table}
 \end{figure*}

A tunable coupling is needed to generate entanglement between two KPOs. Here, we explore a tunable cross-Kerr coupling,
which can be implemented by a nonlinear tunable coupler circuit implemented through a SQUID \cite{kounalakis2018tuneable} as is shown in section \ref{sec:circuit_implementation}.
In anticipation of the effect of a cross-Kerr coupling $K_{12}$, we simultaneously introduce a two-mode drive with strength $\epsilon_{12}$ (see also Fig.~\ref{fig:eigenstates_table}). The total Hamiltonian of the coupled KPOs then becomes
\begin{align}
H=\sum_{j=1}^2 H_{j}
-K_{12}\, a_1^{ \dagger}a_1a_2^{\dagger}a_2 + \epsilon_{12}\, a_1^{ \dagger}a_2^{ \dagger}+ \epsilon_{12}^*\, a_1a_2 \,,
\label{eq:H_full}
\end{align}
where $H_j$ describes the Hamiltonian of the uncoupled KPOs as defined in Eq.~(\ref{eq:H}).

When the KPOs are uncoupled, $K_{12}=\epsilon_{12}=0$, the generalization from the single KPO is trivial: Upon adiabatically turning on $\epsilon_1$ and $\epsilon_2$, a cat state forms in each mode. These two-mode cat states form a degenerate manifold of eigenstates with eigenvalue $\left(|\epsilon_1|^2/K_1+|\epsilon_2|^2/K_2\right)$ spanned by \emph{four} states, which we separate into two groups (see Fig.~\ref{fig:eigenstates_table}): $\ket{C_{\alpha_1}^{\pm},C_{\alpha_2}^{\pm}}$ are even states and $\ket{C_{\alpha_1}^{\pm},C_{\alpha_2}^{\mp}}$ are odd states where the parity of states refers here to the manifolds spanned by two-mode Fock states $\ket{n_1,n_2}$ with $(n_1+n_2)$ being even or odd. Of course, in addition to this total parity, the uncoupled KPOs also conserve parity in the individual modes. Therefore, even states $\ket{C_{\alpha_1}^{\pm},C_{\alpha_2}^{\pm}}$ are adiabatically connected to Fock states $\ket{00}$ and $\ket{11}$, while odd states $\ket{C_{\alpha_1}^{\pm},C_{\alpha_2}^{\mp}}$ are connected to states $\ket{01}$ and $\ket{10}$. Now, in order to create entangled states, such as Bell cat states, via adiabatic evolution, one has to switch on the coupling between the KPOs.

At finite coupling, the total Hamiltonian (\ref{eq:H_full}) can be rewritten in a form equivalent to Eq.~\eqref{eq:H_k}, i.e., 
\begin{align}
H=\sum_{j=1}^2 &\left[ -K_j\left(a_j^{\dagger 2}-\alpha_j^{* 2}\right) \left(a_j^2-\alpha_j^2 \right)+ K_j|\alpha_j|^4\right]\label{eq:H_full_nice}\\
\ &-K_{12} \left(a_1^{ \dagger}a_2^{\dagger}-\frac{\epsilon_{12}^*}{K_{12}}\right)\left(a_1a_2-\frac{\epsilon_{12}}{K_{12}}\right) + \frac{|\epsilon_{12}^2|}{K_{12}},\notag
\end{align}
where $\alpha_j=\sqrt{r_{j}/K_j}e^{i\phi_{j} /2}$ are unaffected by the coupling. We notice that when the coupling parameters are tuned such that $\epsilon_{12}/K_{12} = \alpha_1\alpha_2$, the two-mode coherent states $\ket{\alpha_1,\alpha_2}$ and $\ket{-\alpha_1,-\alpha_2}$ \emph{remain} degenerate eigenstates of the \emph{coupled} Hamiltonian with eigenenergy $K_1|\alpha_1|^4+K_2|\alpha_2|^4+K_{12}|\alpha_1|^2|\alpha_2|^2$. Similarly, if the coupling parameters are tuned such that $\epsilon_{12}/K_{12} = -\alpha_1\alpha_2$, the degenerate subspace is spanned by the odd parity states $\ket{\alpha_1,-\alpha_2}$ and $\ket{-\alpha_1,\alpha_2}$.

Consequently, a finite coupling lifts the degeneracy from four to two degenerate states. This is consistent with the symmetry analysis: While the uncoupled Hamiltonian preserves parity in the two individual modes, the coupled Hamiltonian only preserves the overall parity, i.e., the parity of the \emph{total occupation} of the two modes.

We will now analyze the two specific cases for the coupling parameters in more detail (cf.\ Fig.~\ref{fig:eigenstates_table}) and start with the case $\epsilon_{12}/K_{12} = \alpha_1\alpha_2$. Then, out of the degenerate two-mode coherent states, we can form an even parity degenerate state given by 
\begin{align}
   \ket{ C^+_{\alpha_1 \alpha_2}} =(\ket{\alpha_1,\alpha_2}+\ket{-\alpha_1,- \alpha_2})/N^+_{\alpha_1,\alpha_2}.
\label{eq:even_multimode_state}    
\end{align}
This state connects to the Fock state $\ket{0,0}$ when the drives adiabatically approach zero, i.e. $\epsilon_1=\epsilon_2=\epsilon_{12}\to 0$ .

The complementary odd parity state, degenerate with the one above, reads
\begin{align}
  \ket{C^-_{\alpha_1\alpha_2}} =(\ket{\alpha_1,\alpha_2}-\ket{-\alpha_1,- \alpha_2})/N^-_{\alpha_1,\alpha_2}
\label{eq:odd_multimode_state} \,. 
\end{align}
with $N^\pm_{\alpha_1 \alpha_2}=\sqrt{2(1\pm e^{-2(|\alpha_1|^2+|\alpha_2|^2)})}$.
Towards zero drive, this state smoothly connects to the two degenerate odd parity eigenstates $\ket{0,1}$ and $\ket{1,0}$. Remarkably, starting the adiabatic evolution from zero drive where amplitudes $\alpha_1$ and $\alpha_2$ are small, either state $\ket{0,1}$ or $\ket{1,0}$ or any combination can be prepared initially if the path of drives $\epsilon_1$ and $\epsilon_2$ is correctly adjusted.
This becomes obvious when expanding  $\ket{ C^-_{\alpha_1\alpha_2}}$ for sufficiently small $\alpha_j$ to first order in $\alpha_1$ and $\alpha_2$ which yields
\begin{align}
     \ket{ C^-_{\alpha_1\alpha_2}} \approx \left(\alpha_1\ket{1,0}+\alpha_2\ket{0,1}\right)/\sqrt{|\alpha_1|^2+|\alpha_2|^2}\, .
\end{align}
 When the two displacements now approach zero on a path parametrized by a parameter $\eta$, one finds that the odd parity state $\ket{ C^-_{\alpha_1 \alpha_2}}$ passes over into 
\begin{align}
 \ket{ C^-_{\alpha_1\alpha_2}} \rightarrow \lim_{\eta \rightarrow 0} \frac{\alpha_1(\eta)\ket{1,0}+\alpha_2(\eta)\ket{0,1}}{\sqrt{|\alpha_1(\eta)|^2+|\alpha_2(\eta)|^2}}.
\end{align}
Therefore, when first $\epsilon_2$ ($\epsilon_1$) is turned to zero followed by  $\epsilon_1$ ($\epsilon_2$), the state 
 $\ket{ C^-_{\alpha_1\alpha_2}}$ adiabatically approaches  $\ket{1,0}$ ($\ket{0,1}$).

The main result of this analysis is that the even $\ket{ C^+_{\alpha_1\alpha_2}}$ and odd $\ket{ C^-_{\alpha_1\alpha_2}}$ parity states can be prepared by following an adiabatic evolution from respective Fock states. The significance of these states is that they are entangled two-mode coherent states. As we will show below, in the limit of sufficiently large $\alpha_1,\alpha_2\gtrsim 1$, these states  become excellent approximations for two different two-mode Bell cat states.
For this reason, we dub the states $\ket{ C^+_{\alpha_1 \alpha_2}}$ and $\ket{ C^-_{\alpha_1 \alpha_2}}$ as \textit{proto-Bell-cat states}.

The above protocol can be used to generate two of the four proto-Bell-cat states. The other two can be obtained by performing the adiabatic evolution under the other constraint $\epsilon_{12}/K_{12} = -\alpha_1\alpha_2$, with opposite sign as before. The analysis is completely analogous. In this case, the even and odd parity degenerate states are given by
\begin{align}
   \ket{ C^\pm_{\alpha_1\ -\alpha_2}} =(\ket{\alpha_1,-\alpha_2}\pm\ket{-\alpha_1,\alpha_2})/N^\pm_{\alpha_1\ -\alpha_2}.
\label{eq:both_proto_Bell_state}    
\end{align}
The even parity state $\ket{ C^+_{\alpha_1\ -\alpha_2}}$ connects adiabatically to the Fock state $\ket{00}$, while the odd parity state $\ket{ C^-_{\alpha_1\ -\alpha_2}}$ approaches
\begin{align}
 \ket{ C^-_{\alpha_1\ -\alpha_2}} \rightarrow \lim_{\eta \rightarrow 0} \frac{\alpha_1(\eta)\ket{1,0}-\alpha_2(\eta)\ket{0,1}}{\sqrt{|\alpha_1(\eta)|^2+|\alpha_2(\eta)|^2}}.
\end{align}
A summary of the eigenstates of the two-mode Hamiltonians is provided in Fig.~\ref{fig:eigenstates_table}.

\subsection{ Initialization from Fock states to Bell cat states}
\label{sec:FocktoBell}

In the previous section, we demonstrated the adiabatic connection between Fock states and proto-Bell-cat states $ \ket{ C^\pm_{\alpha_1\ \pm\alpha_2}}$. Here, we aim to extend the adiabatic connection to the four true Bell cat states defined by
\begin {align}
\ket{\Phi^\pm_{\alpha_1 \alpha_2}}=&\frac{1}{\sqrt{2}}\left(\ket{C_{\alpha_1}^+,C_{\alpha_2}^+}\pm\ket{C_{\alpha_1}^-,C_{\alpha_2}^-}\right) ,\\ 
\ket{\Psi^\pm_{\alpha_1 \alpha_2}}=&\frac{1}{\sqrt{2}}\left(\ket{C_{\alpha_1}^+,C_{\alpha_2}^-}\pm\ket{C_{\alpha_1}^-,C_{\alpha_2}^+}\right)\nonumber.
\end{align}
These Bell cat states are maximally entangled states defined in the computational space of the two cat-qubits. According to their composition out of two-mode coherent states, the Bell states $\ket{\Phi^\pm_{\alpha_1 \alpha_2}}$ have even total parity, while states $\ket{\Psi^\pm_{\alpha_1 \alpha_2}}$ are odd.

 \begin{figure*}
		
        \includegraphics[width=0.7\linewidth]{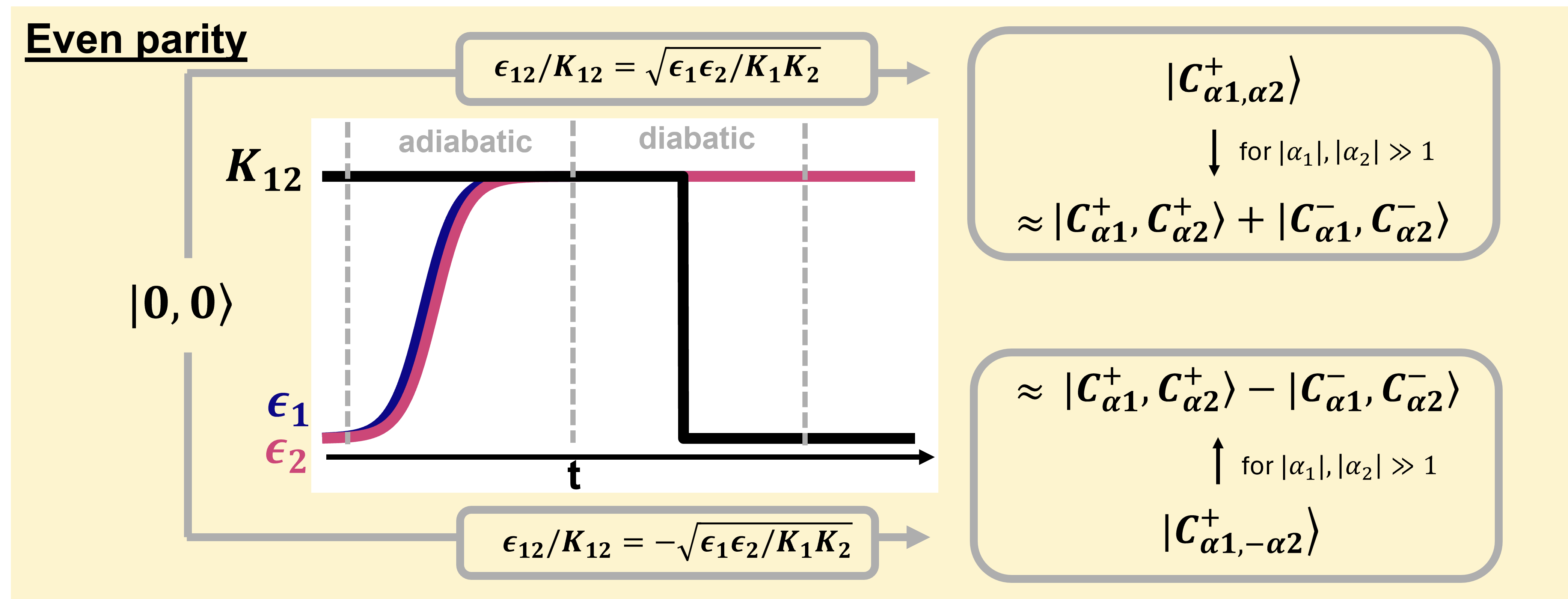}
          \includegraphics[width=0.7\linewidth]{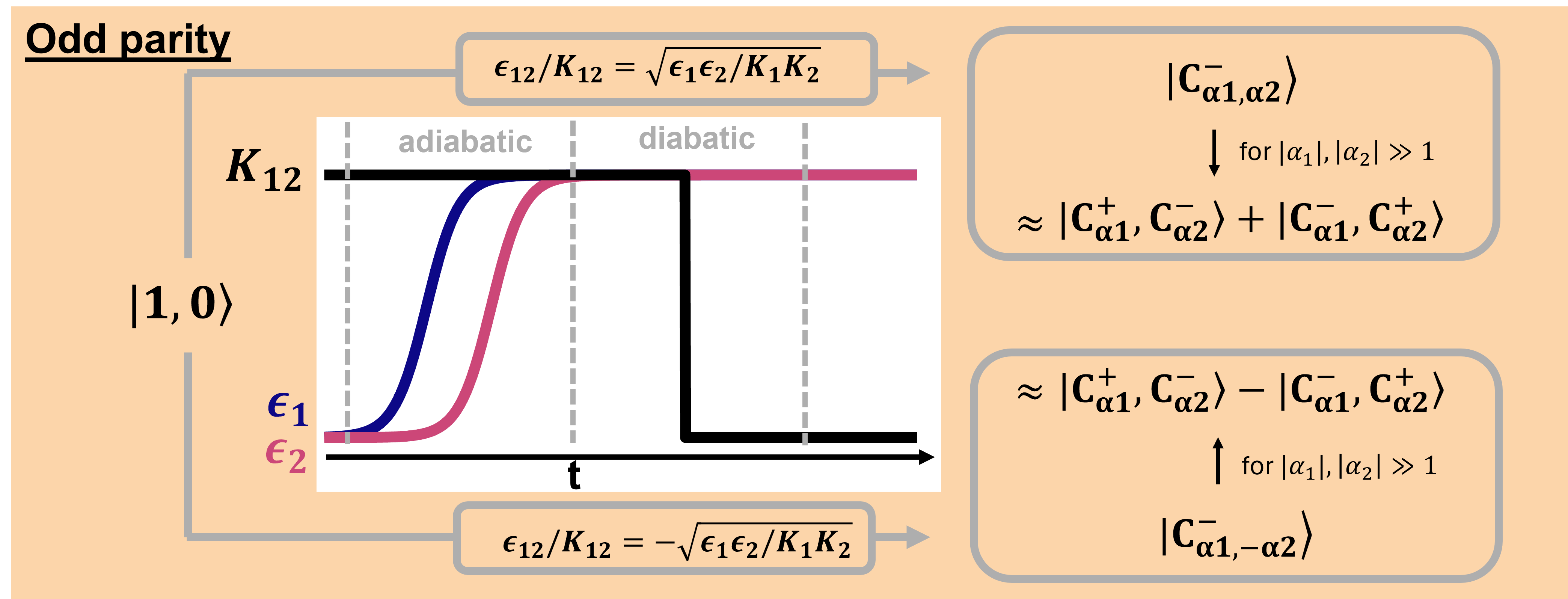}
	
	\caption{
    Drives and cross-Kerr coupling as a function of time that achieve the adiabatic connection between Fock states $\ket{0,0}$ and $\ket{1,0}$ and corresponding Bell cat states. The two-mode drive $\epsilon_{12}$ must be tuned according to the relations in the corresponding gray boxes at all times. 
    }
\label{fig:transitions}
 \end{figure*}
 
To establish the direct relation between the proto-Bell states in \cref{eq:even_multimode_state,eq:odd_multimode_state,eq:both_proto_Bell_state} and the true Bell cat states, it is useful to rewrite the proto-Bell states into the computational basis of two-qubit cat states spanned by $\ket{C_{\alpha_1}^\pm,C_{\alpha_2}^\pm}$ and $\ket{C_{\alpha_1}^\pm,C_{\alpha_2}^\mp}$. In this basis, the proto-Bell states are exact superpositions of the two Bell states with concurrent parity, e.g. for the even parity state
\begin{align}
\ket{ C^+_{\alpha_1 \alpha_2}} =\nu_+ \ket{\Phi^+_{\alpha_1 \alpha_2}}+\nu_-\ket{\Phi^-_{\alpha_1 \alpha_2}}.
\end{align}
The amplitudes
\[ \nu_\pm=(N_{\alpha_1}^+N_{\alpha_2}^+ \pm N_{\alpha_1}^-N_{\alpha_2}^-)/(2\sqrt{2}N^+_{\alpha_1 \alpha_2})\] contain normalization factors for single mode cat states $N_{\alpha_i}$ defined in Sec.~\ref{sec:single_KPO} and two-mode proto-Bell-cat states $N^+_{\alpha_1 \alpha_2}$ defined in Sec.~\ref{sec:coupled_KPOs}.

Now, the important point is that for sufficiently large driving strength, i.e.\ in the limit of large amplitudes $\alpha_1,\alpha_2\gg 1$, the single mode normalizations almost coincide $N^+_{\alpha_i} \simeq N^-_{\alpha_i}$ so that $\nu_-$ is exponentially suppressed and the proto-Bell-cat state $\ket{ C^+_{\alpha_1,\alpha_2}}$ becomes an excellent approximation for the true Bell cat state $\ket{\Phi^+_{\alpha_1,\alpha_2}}$.
Likewise, one verifies that this applies to all four Bell cat states which leads to the crucial relations
\begin{align}
\ket{\Phi^+_{\alpha_1 \alpha_2}}\simeq \ket{C^+_{\alpha_1 \alpha_2}},\quad \ket{\Phi^-_{\alpha_1 \alpha_2}}\simeq \ket{C^+_{\alpha_1\ -\alpha_2}},\nonumber\\ 
\ket{\Psi^+_{\alpha_1 \alpha_2}}\simeq \ket{C^-_{\alpha_1 \alpha_2}},\quad \ket{\Psi^-_{\alpha_1 \alpha_2}}\simeq \ket{C^-_{\alpha_1\ -\alpha_2}}.\nonumber
\end{align}

To summarize, the protocol to prepare two cat qubits into either of the four Bell cat states is as follows, see Fig.~\ref{fig:transitions}: In a first step, the system has to be initialized to a Fock state with total parity equal to that of the desired Bell state, e.g. $\ket{0,0}$ for even total parity and $\ket{0,1}$ for odd total parity. Diabatically switching on the cross-Kerr interaction will leave this state unchanged. In a second step, the two-photon drives are switched on adiabatically along either of the constraints $\epsilon_{12}=\pm K_{12}\, \alpha_1\alpha_2$, where the sign determines the final Bell cat state. This protocol leads to the proto-Bell-cat states. Then, by adiabatically increasing the amplitudes $\alpha_1$ and $\alpha_2$, the proto-Bell-cat states converge towards the true Bell cat states. Notably, as we will show below, $\alpha_1$ and $\alpha_2$ need to be tuned only slightly above unity to reach the desired Bell cat state with excellent fidelity. Finally, if convenient, the cross-Kerr interaction as well as the two-mode drive can be turned off diabatically which does not affect the Bell cat states since they are eigenstates of both the interacting Hamiltonian and the non-interacting one.

\section{Geometric transitions between Bell cat states}
\label{sec:geometric}

\begin{figure}	
	\begin{center}
		\includegraphics[width=\linewidth]{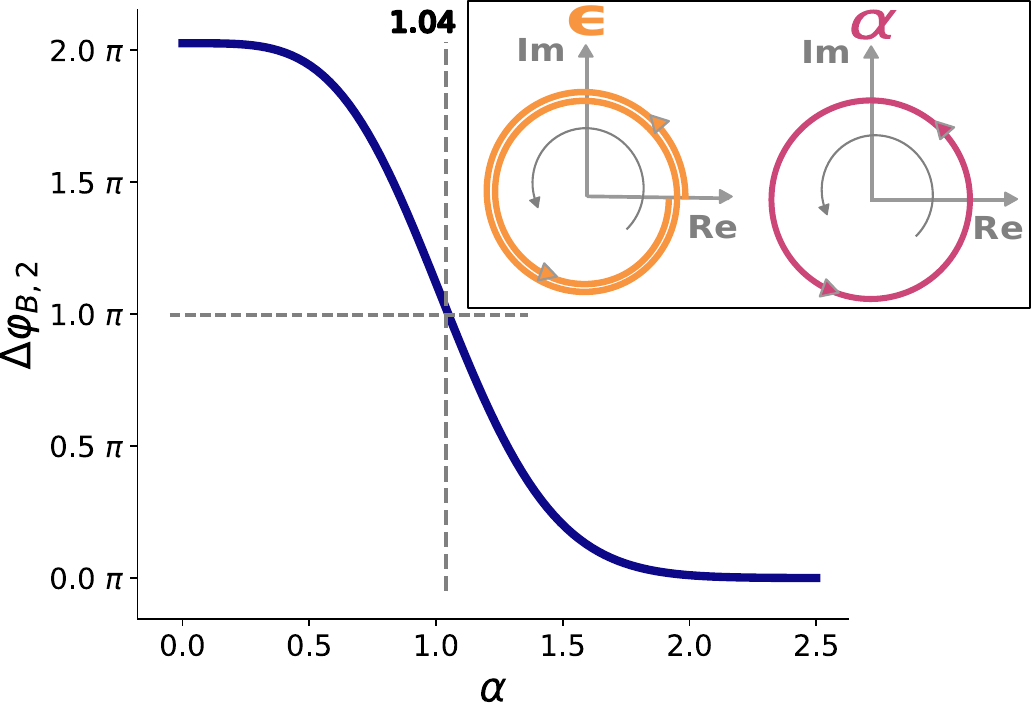}
	\end{center}
	\caption{Difference between Berry phases accumulated by states $\ket{C^+_{\alpha}}$ and $\ket{C^-_{\alpha}}$ after two rotations of the drive amplitude $\epsilon$. The dashed lines indicate a phase difference of $\pi$.}
\label{fig:phase_difference_alpha}
 \end{figure}

Since the two pairs of Bell states each have a defined total parity, it is also possible to directly transform between individual Bell states within a pair with given total parity. 
Thereby, we can make use of the fact that 
degenerate cat states with different parity accumulate a difference in geometric phase under adiabatic variations of their drives  \cite{puri2020bias,wang2023scheme,kang2022nonadiabatic} while their dynamical phases are equal. To realize a transition between two Bell states within the same parity sector, at least one of the two-mode drives with complex amplitude $\epsilon_j$ has to be adiabatically rotated on a closed path around the origin. This procedure generates a Berry phase which does depend on the chosen path of rotation. 
For a path $\mathcal{C}$ in the complex plane of amplitude $\epsilon_j$, the accumulated geometrical phase of a cat state with even/odd parity and displacement $\alpha_j$ turns out to be (see \cite{supplementary_information}), 
\begin{align}
 \varphi_B^\pm = - \int_\mathcal{C}  d\phi_j\ \frac{|\alpha_j|^2}{2} \left[  \frac{1\mp e^{-2|\alpha_j|^2}}{1\pm e^{-2|\alpha_j|^2}}\right].
\end{align}
Apparently, this Berry phase is finite only for paths along which the phase $\phi_j$ of the two-photon drive is varied.
Crucially, the Berry phases accumulated by even and odd cat states differ. The acquired geometrical phase can be traced back to the non-zero overlap between coherent states $\ket{\alpha_j}$ and $\ket{-\alpha_j}$. As this overlap vanishes for $|\alpha_j|\gg 1$, so does the difference in Berry phase.
Considering a specific closed path $\mathcal{C}$ where the drive adiabatically rotates around the origin with constant $|\alpha_j|$, the difference in Berry phase for one rotation is given by
\begin{align}
 \Delta \varphi_{B,1} =-\pi|\alpha_j|^2 \left[  \frac{ -4 e^{-2|\alpha_j|^2}}{1-e^{-4|\alpha_j|^2}}\right]  .
\end{align}
It is important to note that during a rotation of the drive around the origin the even and odd cat states additionally acquire a "parity phase" difference due to the different periodicities of $\ket{C^+_{\alpha j}}$ and $\ket{C^-_{\alpha j}}$ as functions of the phase of the two-photon drive, i.e.

\begin{align}
\ket{C^+_{\alpha_j}}\underset{\strut \phi_{j}=0\ \leftrightarrow \ 2\pi}{\strut \longleftrightarrow} +\ket{C^+_{\alpha_j}}\\
\ket{C^-_{\alpha_j}}\overset{\strut \alpha_j\ \leftrightarrow\ -\alpha_j}{\strut \longleftrightarrow} -\ket{C^-_{\alpha_j}}. \nonumber
\end{align}
\\
While the "parity phase" can also be used for transitions between states with a fixed phase difference $\pi$ \cite{puri2020bias},  for the purpose of exploiting the geometric transition we now consider the case where the drive is rotated around the origin twice, resulting in no parity phase difference between even and odd states and only a difference in Berry phase of $\Delta \varphi_{B,2} = 2 \Delta \varphi_{B,1} $. This difference is illustrated  in Fig.~\ref{fig:phase_difference_alpha}. Obviously, within the range $0.5\lesssim \alpha_j \lesssim 1.5$ the rotation in drive space indeed induces a substantial phase difference on the order of $\pi$.
 
In general, a state initialized in $\ket{\Psi_0}=\gamma_+ \ket{C_\alpha^+}\pm \gamma_- \ket{C_\alpha^-}$ will after $2M$ rotations up to a global phase arrive at
\begin{align}
\ket{\Psi_{2M}} =  \gamma_+ \ket{C_{\alpha}^{+}} \pm e^{-i \Delta \varphi_{B,2M}(\alpha)} \gamma_- \ket{C_{\alpha}^{-}} ,
\label{eq:result_state}
\end{align}
with $ \Delta \varphi_{B,2M}(\alpha) =  2M\Delta \varphi_{B,1}(\alpha)$.
Likewise, a two-mode state initialized in $\gamma_+ \ket{C_{\alpha_1}^+,C_{\alpha_2}^+}\pm \gamma_- \ket{C_{\alpha_1}^\shortminus,C_{\alpha_2}^\shortminus}$
will be after $2M$ rotations of the drive of mode $j$ in state 
\begin{align}
  \ket{\Psi_{2M}} =   \gamma_+ \ket{C_{\alpha_1}^+,C_{\alpha_2}^+}\pm e^{-i \Delta \varphi_{B,2M}(\alpha_j)}\gamma_- \ket{C_{\alpha_1}^\shortminus,C_{\alpha_2}^\shortminus}.
  \label{eq:state_multimode_berry_phase}
\end{align}
To generate an arbitrary phase between superpositions of cat states, it is always possible to define a trajectory in the space of the phase of an individual drive which provides the desired difference in Berry phases.
Another option is to increase the accumulated Berry phase by rotating both drives simultaneously, thus generating a phase difference that equals the sum of the differences of the corresponding Berry phases.

\section{Multi-mode entangled states}
\label{sec:multimode}

The scheme developed in Sec.~\ref{sec:cat_qubit} easily scales up to generate entanglement between several Kerr parametric oscillators.
For this purpose, we now consider a system of $N$ different KPOs where neighbors are coupled via tunable cross-Kerr $K_{j,j+1}$ and mode-mixing terms $\epsilon_{j,j+1}$, which generalize the system Hamiltonian in Eq.~\eqref{eq:H_full_nice}.

Now we assume that the first $N-1$ modes of the system are already prepared in a multi-mode cat state. 
For the first two modes, the corresponding protocol has been described above in Sec.~\ref{sec:FocktoBell}. 
Mode $N$ is initialized in its ground state (Fock state) so that the total initial state of the system is given by
\begin{align}
   \ket{\Psi_0} =\frac{1}{N^\pm(\vec{\tilde{\alpha}})} \left( \ket{\alpha_1,...,\alpha_{N-1}}\pm \ket{\shortminus\alpha_1,...,\shortminus\alpha_{N-1}}\right)\otimes \ket{0},
\end{align}
with $N^\pm(\vec{\tilde{\alpha}}) $ being the normalization.
This state is an eigenstate of the uncoupled system Hamiltonian, where all cross-Kerr terms $K_{j\, j+1}$ and all mode-mixing drives $\epsilon_{j\, j+1}$ are turned off. The protocol for all modes up to $N-1$ gives rise to coherent state amplitudes $\alpha_j=\sqrt{r_{j}/K_j}e^{i\phi_{j} /2}$, $j=1, \ldots N-1$. Up to this point, the $N$th mode is uncoupled and not driven $\epsilon_{N} = 0$.
Now, diabatically tuning the cross-Kerr coupling $K_{N-1\,  N}$ between mode $N-1$ and mode $N$ to a finite value leaves the total state unchanged, as it remains an eigenstate of the coupled system. Further, the mode-mixing drive $\epsilon_{N-1\, N}$ and two-photon drive $\epsilon_N$ are turned on adiabatically under the constraint  $\epsilon_{N-1\, N}/K_{N-1\, N} = \alpha_{N-1}\alpha_{N}$, with $\alpha_N=\sqrt{r_N/K_N} \, e^{i\displaystyle\phi_{N} /2}$. Via this adiabatic drive the state $\ket{\Psi_0}$ changes into the multi-mode cat state
\begin{align}
\ket{C_{\vec{\alpha}}}  =1/N^\pm_{\vec{\alpha}} ( \ket{\alpha_1,...,\alpha_{N}}\pm \ket{\shortminus\alpha_1,...,\shortminus\alpha_{N}}).
\end{align}
Successively applying this protocol and selecting $\epsilon_{j\, j+1}/K_{j\,j+1} = \pm \alpha_{j}\alpha_{j+1}$ in each step, bridges adiabatically the Fock states $\Motimes_{j=1}^N \ket{0}_j$ and  $\ket{1}\Motimes_{j=2}^N \ket{0}_j$ with any multi-mode cat state $C^\pm(\vec{\alpha},\vec{\sigma})$ defined by
\begin{align}
C^\pm(\vec{\alpha},\vec{\sigma}) = \frac{1}{N^\pm(\vec{\alpha},\vec{\sigma})} \left( \Motimes_{j=1}^N \ket{\sigma_j\alpha_j}\pm \Motimes_{j=1}^N \ket{\shortminus\sigma_j\alpha_j}\right),
\end{align}
with normalization $N^\pm(\vec{\alpha},\vec{\sigma})$  and sign   $ \sigma_j = \pm1$.

\section{Feasibility}
We now discuss how such an interaction can be implemented in superconducting circuits and how limitations in the accuracy of the individual steps influence the fidelity of the final state.
\subsection{Circuit implementation}
\label{sec:circuit_implementation}
\begin{figure}	
	\begin{center}
		\includegraphics[width=.7\linewidth]{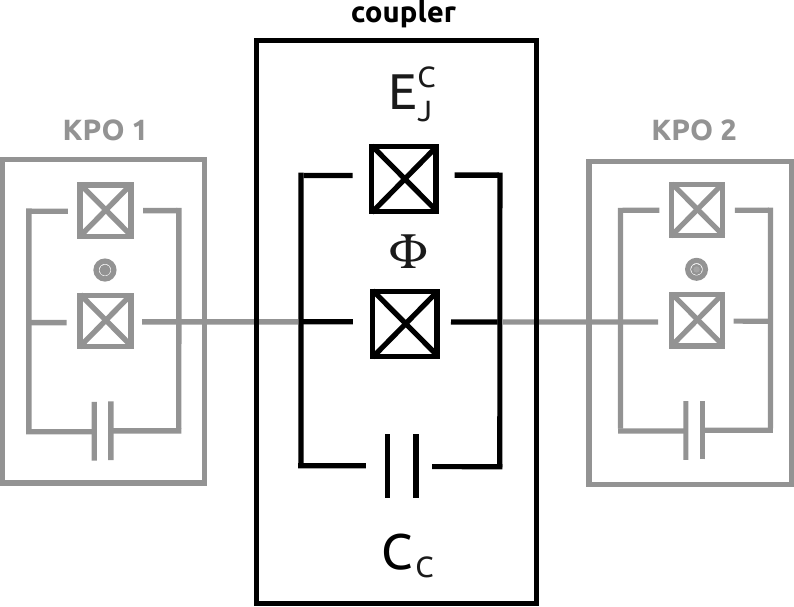}
	\end{center}
	\caption{Layout of a possible physical implementation of the coupling via two transmons that are connected by a tunable coupler. The transmons as well as the coupler circuit are realized using SQUIDs. }
\label{fig:layout_circuit}
 \end{figure}

An efficient way to implement a tunable cross Kerr coupling between two neighboring cat qubits is by using a nonlinear tunable coupler circuit. One of the simplest examples is to couple each pair of KPOs by a SQUID circuit characterized by its flux-tunable Josephson energy $E^C_J(\Phi)$ and a capacitance $C_C$, see Fig.~\ref{fig:layout_circuit}.

The nonlinearity of the SQUID coupler circuit gives rise to nonlinear interactions between the two modes, which can be tuned during the protocol. The leading order non-linearity of the circuit in Fig.~\ref{fig:layout_circuit} has been worked out analytically in Ref.~\cite{kounalakis2018tuneable}. In the rotating frame of each KPO, the Hamiltonian of the circuit is given by
\begin{align}
    H/\hbar = \sum_{i=1}^2 -K_i a_i^{\dagger2}a_i^2 
    - K_{12}a_1^\dagger a_1 a_2^\dagger a_2,
\end{align}
The cross Kerr coupling is given by $K_{12} = E_J^C (\delta_1\delta_2)^2/16$, where $\delta_i$ is the zero point fluctuation corresponding to mode $i$, that depends on the implementation of the KPOs. The strength of the individual Kerr terms is independent of the inductance of the coupler. This allows switching the cross Kerr interaction on and off without affecting the individual Kerr terms

As pointed out in Ref.~\cite{kounalakis2018tuneable}, an undesirable linear coupling term $J(a_1^\dagger a_2+a_2^\dagger a_1)$ is present in the circuit. The effect of this term is minimized when the two KPOs are asymmetric, i.e. $K_1$ and $K_2$ are incommensurate, such that any unwanted coupling can be neglected within the RWA. The required asymmetry of the KPOs is fully compatible with our proposed initialization method.

\subsection{Robustness}

So far we have assumed ideal conditions for the protocol displayed in Fig.~\ref{fig:transitions} which implies an adiabatic ramp-up of the driving strengths $\epsilon_1, \epsilon_2$ to approach  the resonance conditions $\epsilon_{12}/K_{12}=\pm \sqrt{\epsilon_1\epsilon_2/K_1 K_2}$ and a subsequent diabatic switching-off of both $\epsilon_{12}, K_{12}$. Here, we discuss the feasibility of this protocol under realistic conditions. 

For this purpose, 
 we start by discussing the fidelity of the transition from individual Fock states at $t=0$ to entangled Bell cat states within a finite time interval $\tau$. The crucial questions are then how fast the drive amplitudes can be changed while preserving the main features of the adiabatic/diabatic evolution. 

\begin{figure}	
	\begin{center}
	\includegraphics[width=\linewidth]{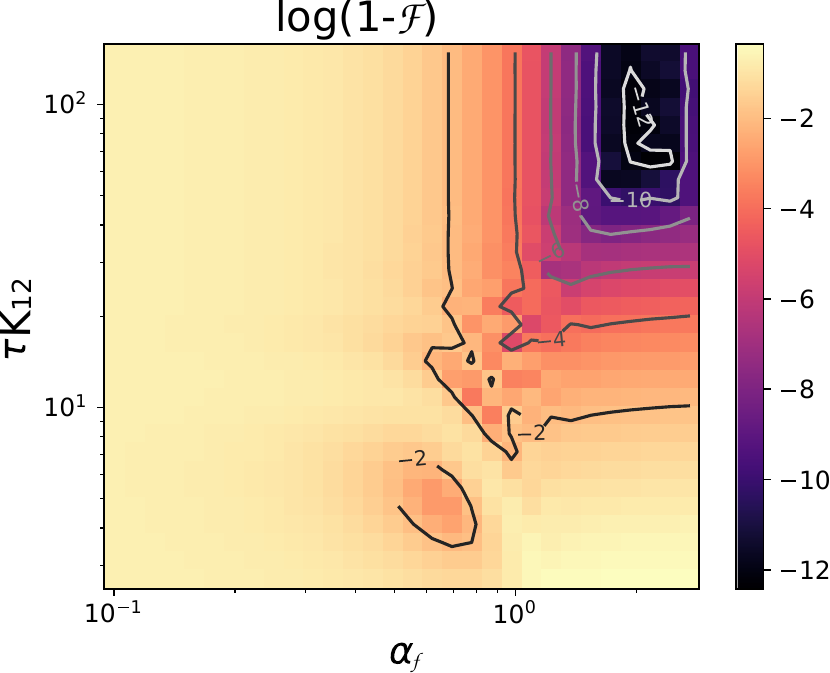}
	\end{center}
	\caption{Simulation of the transformation to Bell state $\ket{\Phi^+_{\alpha_f,\alpha_f}}$ from Fock state $\ket{0,0}$. Plotted is the logarithmic infidelity of the final state as a function of displacements $|\alpha_1| = |\alpha_2| \equiv \alpha_f$ and of the scaled ramp-up time $\tau K_{12}$, where the ramping of the drive is $\epsilon_1(t)= \epsilon_2(t)\propto \tanh{(4 (2 t/\tau-1))}+1$. } 
\label{fig:fidelity_tau}
 \end{figure}

 Let us first consider the adiabatic segment of the protocol. This depends on the energy difference between first excited and ground state within a given parity manifold and therefore on the strength of the Kerr nonlinearities.
 More specifically, we consider the system being initialized in the Fock state $|\psi(t=0)\rangle=\ket{0,0}$ and then follow numerically the time evolution $|\psi(t)\rangle$ when the  drive amplitudes $\epsilon_1(t)=\epsilon_2(t)\equiv \epsilon(t)$ are smoothly switched-on  $\epsilon(t) \propto \tanh{(8 t/\tau-4)}+1 $. For simplicity we set $K_{12}=K_1=K_2$ and choose as a target state $|\Phi_{\alpha_1\, \alpha_2}^+\rangle $ with identical final displacements 
 $\alpha_{1, \textrm{final}} = \alpha_{2, \textrm{final}} \equiv \alpha_f>0$. The corresponding fidelity $
 \mathcal{F}=|\langle \psi(t)|\Phi^+_{\alpha_f\, \alpha_f}\rangle|^2$ is displayed in Fig. \ref{fig:fidelity_tau} as a function of the drive-ramp duration $\tau$ and the final displacement $\alpha_f$.
As expected, the targeted Bell state is reached with high fidelity for a sufficiently long ramp time $\tau$ and a sufficiently large final amplitude $\alpha_f$. Notably though, we find a fidelity close to 1 ($\mathcal{F}>0.9999$) already for finite ramp duration with $K_{12}\tau \gtrsim 20 $ and $\alpha_f \gtrsim 1$. In terms of physical parameters this corresponds for typical values of the Kerr nonlinearity in the MHz regime to ramp times on the order of $\mu s$. 

\begin{figure}
	\begin{center}
	\includegraphics[width=\linewidth]{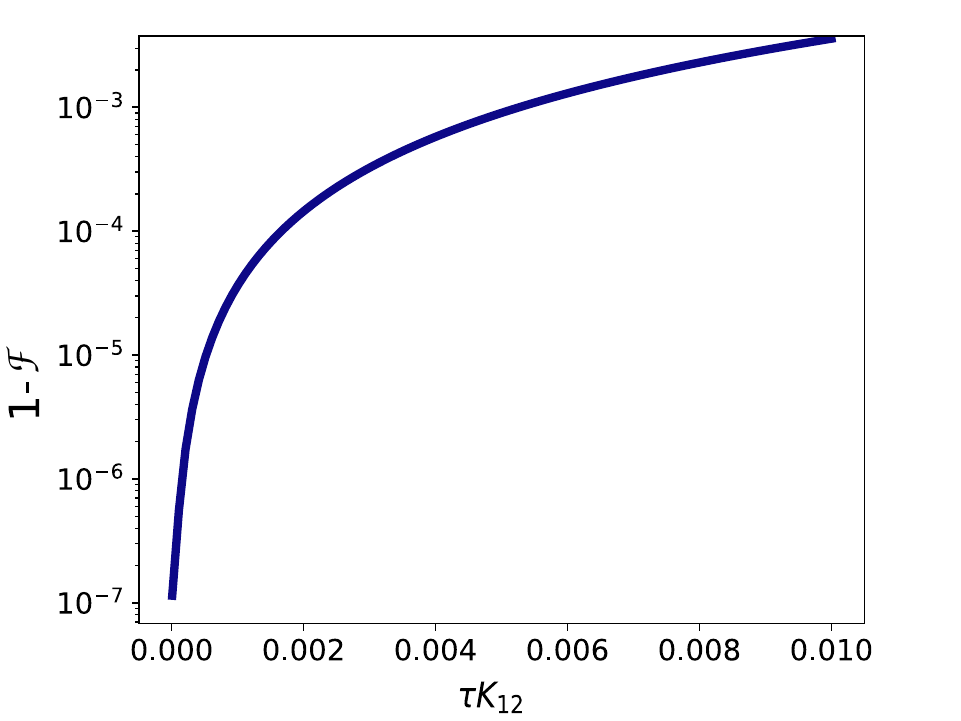}
	\end{center}
    \caption{Fidelity of a state initialized in $\ket{\Phi^+_{\alpha_1,\alpha_2}}$ after switching off the interaction between the modes. The mode mixing drive $\epsilon_{12}$ is assumed to be switched off instantaneously while the cross-Kerr interaction is assumed to be switched off within a finite time $\tau$. The amplitudes are $\alpha_1 = \alpha_2 \approx 2$.}
\label{fig:fidelity_deviation_k12}
\end{figure}
In a second step, we investigate how susceptible our protocol is to deviations from the condition $\epsilon_{12} = \pm K_{12}\alpha_1\alpha_2$ due to experimental constraints. As an example, we assume that the two-mode drive $\epsilon_{12}$ can be switched off instantaneously while switching off the cross-Kerr interaction $K_{12}$ takes a finite amount of time, thus violating the required relation in the diabatic segment of the protocol, see Fig. \ref{fig:transitions}.
To investigate the effect of such a violation, we analyze the fidelity of the final state with respect to the initial state since in our protocol the state is assumed to stay the same during this transition.
As seen in  Fig. \ref{fig:fidelity_deviation_k12},  as long as the cross-Kerr interaction can be turned off fast enough, i.e. $K_{12}\tau\lesssim 0.002$ (corresponding to time spans in the ns domain), the final state will still be reached with fidelity $\mathcal{F}>0.999$.

Finally, we address protocols for the manipulation of Bell cats via geometric transitions: One mode $j$ is rotated by tuning  drive amplitudes in a manner that results in a difference of $\pi$ in the Berry phase, and thus in a sign change in the superposition of states constituting a given Bell state. For this purpose, it is helpful to first consider the single-mode case of a single cat qubit. There, the rotation corresponds to a transition from $\ket{\alpha}$ to $\ket{-\alpha}$, where a geometric phase of $\pi$ appears after two adiabatic rotations of the drive if
\begin{align}
  \pi \stackrel{!}{=}\Delta \varphi_{B,2} &=- 2\pi |\alpha|^2 \left[  \frac{ -4 e^{-2|\alpha|^2}}{1-e^{-4|\alpha|^2}}\right]   
\end{align}
which implies $
\sinh(2|\alpha|^2) = 4 |\alpha|^2 \ \Rightarrow \ |\alpha|\approx 1.04 $. One can easily check 
that after two rotations in the parameter space of the complex drive amplitude the initial state $\ket{\alpha =1.04}$ ends up as $\ket{\alpha=-1.04}$, while following the same protocol at large $|\alpha| \gg 1$ leaves the initial state basically unchanged (see \cite{supplementary_information}). 

 \begin{figure*}
	\begin{center}
	\includegraphics[width=0.8\linewidth]{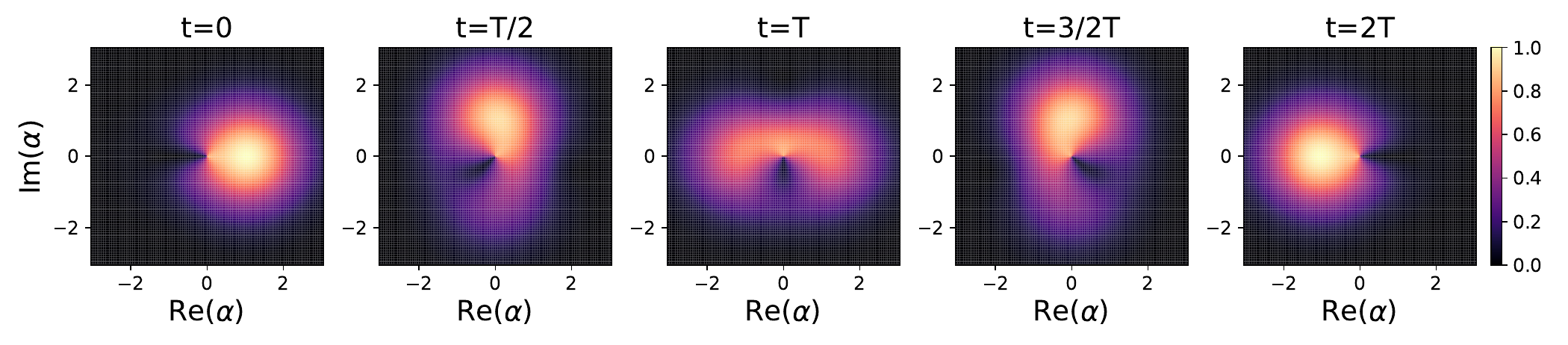}
    \includegraphics[width=0.75\linewidth]{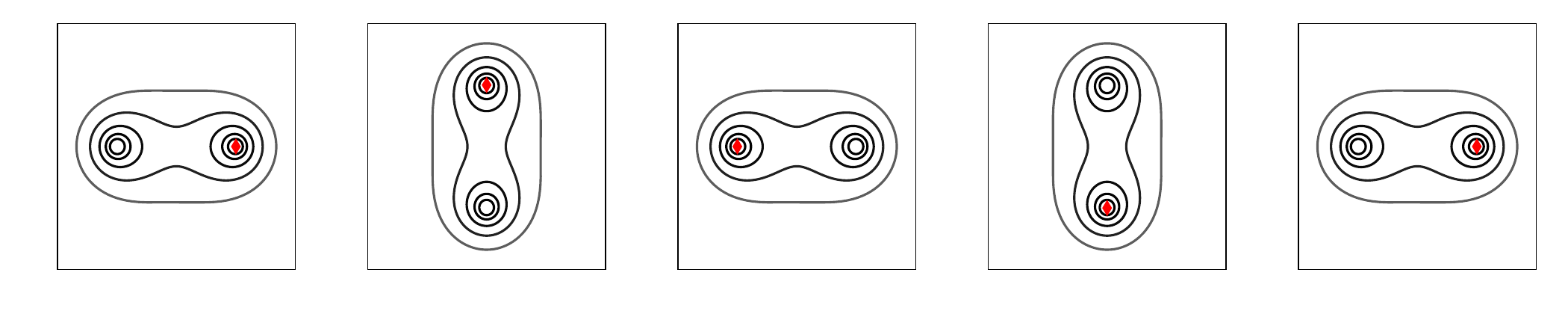}
    \includegraphics[width=0.8\linewidth]{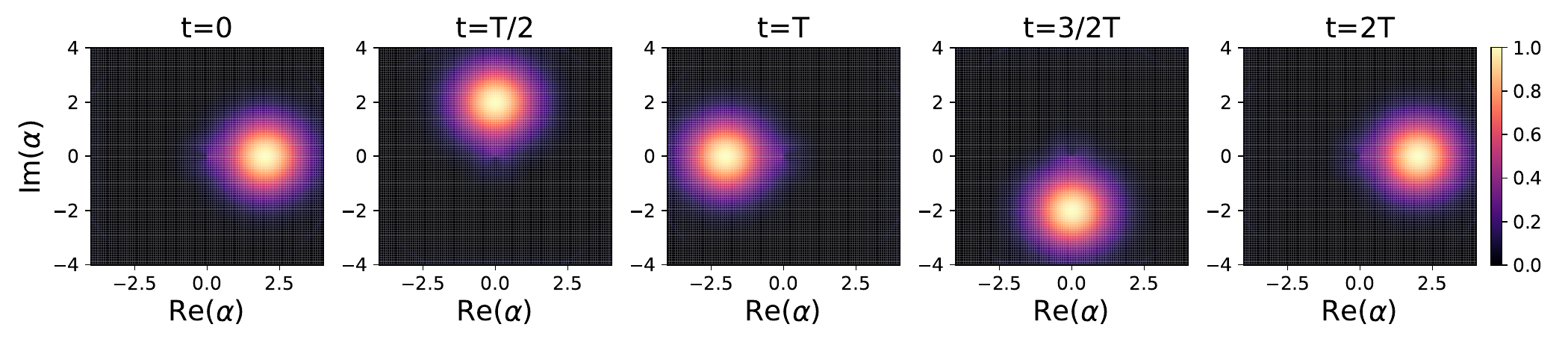}
	\end{center}
	\caption{Evolution of the state of a system that was initialized in $\ket{\Psi(0)}=\ket{\Phi^+_{\alpha_1,\alpha_2}}$. The simulation is done for $\alpha=1.04$ (top) and $\alpha=2$ (bottom). Plotted is a projection onto the two Bell states with even parity, where $\alpha_2$ is fixed to the value of the initial state of the system while $\alpha_1$ is changed depending on $\alpha$. The image shows the projection $|\braket{\Phi^+_{\alpha,\alpha_2}|\Psi(t)}|$ for $Re(\alpha)>0$ and $|\braket{\Phi^-_{\shortminus\alpha,\alpha_2}|\Psi(t)}|$ for $Re(\alpha)<0$. The middle image shows a contour plot of the potential of the system where the red diamond indicates the potential well that the system was initialized in.}
\label{fig:transition_projection_Bell}
\end{figure*}

Let us now turn to the two-mode cate states. We consider a situation where the transition from one Bell state with even total parity, $|\psi(t=0)\rangle=\ket{\Phi^+_{\alpha_1 \, \alpha_2}}$,   to the complementary Bell state with even total parity $\ket{\Phi^-_{\alpha_1,\alpha_2}}$ is realized, see Eq.\eqref{eq:state_multimode_berry_phase} and Fig. \ref{fig:transition_projection_Bell}. 
A convenient way to visualize the evolution of this multi-mode state $|\Psi(t)\rangle$ in the four dimensional phase space is to project it  at various instances in time onto Bell states with even parity $|\Phi_{\alpha_1\, \alpha_2}^\pm\rangle$ at a fixed  phase-space variable $\alpha_2$ and a varying $\alpha_1=\alpha$, see Fig. \ref{fig:transition_projection_Bell}. To be more precise, we show the projection $|\braket{\Phi^+_{\alpha,\alpha_2}|\Psi(t)}|$ for $\mathrm{Re}(\alpha)>0$ and $|\braket{\Phi^-_{\shortminus\alpha,\alpha_2}|\Psi(t)}|$ for $\mathrm{Re}(\alpha)<0$, thus mimicking a Husimi-Q function on a two-dimensional cut of phase-space. The appealing feature of this presentation becomes apparent when considering the evolution for a large displacement $\alpha = 2$, see lower row of Fig. \ref{fig:transition_projection_Bell}, where  the state simply follows the full rotation of $\alpha$ without acquiring a substantial geometric phase. Accordingly,  the system evolves from $\ket{\Phi^+_{\alpha\, \alpha_2}}$  and returns to this state.  For $\alpha_1 =1.04 $, however, the initial state $\ket{\Phi^+_{\alpha\,\alpha_2}}$ is indeed transformed  to $\ket{\Phi^-_{\alpha\, \alpha_2}}$ due to an emerging Berry phase of $\pi$. The fidelity of $\mathcal{F}>0.9999$ indicates that the tuning of the drive was slow enough to be considered as being adiabatic.

\section{Conclusion}
Quantum information processing with continuous variables offers the prospect of inherent protection against certain types of noise. As a necessary pre-requisite it requires a robust and scalable protocol for the creation and manipulation of entangled cat states, so-called Bell cats. In this work we have developed a protocol which fulfills these requirements. It starts from individual Fock states and connects them to Bell cats while always staying in the protected ground state of the system. A tunable cross-Kerr coupling between two Kerr parametric oscillators accompanied by a tunable two-mode drive is sufficient for the creation of all four Bell cats that arise  naturally as ground states of the interacting Hamiltonian. The protocol is relatively simple and consists of one adiabatic and one diabatic segment together with matching conditions between the drive parameters and the Kerr coefficients. Important for practical realizations is the fact that this protocol is robust against finite-time implementations as we demonstrate explicitly. Once Bell cats are created, they can be manipulated by exploiting geometric phases that emerge when drive amplitudes are tuned in closed loops in parameter space. A notable benefit of the protocol is its scalability to multi-mode Bell cats which allows  for the creation of complex quantum states for information processing and simulations. While parameters that we discuss in this work apply particularly to superconducting circuits, for example, in form of transmon-type realizations, the scheme can easily  be adapted to other platforms as well.

\section{Acknowledgments}
We gratefully acknowledge financial support of the Baden-Württemberg Stiftung through the Quantum Technology Network and of the German Science Foundation (DFG) through AN336/18-1. The authors would also like to thank I. M. Pop for the fruitful discussions.

\section{Author contributions}
M.R. performed the calculations and developed the simulations. C.P., B.K., and J.A. supervised the work. M.R. wrote the manuscript and C.P., B.K., and J.A. edited the manuscript.
All authors contributed to the discussions and interpretations of the results.

\section{Competing interests}
The authors declare that they have no competing interests.

\bibliography{lit}
\newpage

\onecolumngrid 
\newpage

\pagebreak

\begin{center}
\textbf{\Large Supplementary information to "Fast initialization of Bell states with Schrödinger cats in  multi-mode systems" }
\end{center}

\setcounter{equation}{0}
\setcounter{section}{0}

\setcounter{figure}{0}
\setcounter{table}{0}
\setcounter{page}{1}
\makeatletter
\renewcommand{\theequation}{S\arabic{equation}}
\renewcommand{\thefigure}{S\arabic{figure}}
\renewcommand{\thesection}{S\arabic{section}}
\renewcommand{\bibnumfmt}[1]{[S#1]}
\renewcommand{\citenumfont}[1]{S#1}

\section{Connection between Fock states and odd parity cat states}
To analyze the connection between odd parity Fock states and cat states of equal parity we calculate $ \lim_{\eta \rightarrow 0}\ket{ C^-_{\alpha_1(\eta),\alpha_2(\eta)}} $ where both $\alpha_1$ and $\alpha_2$ go to 0 for $\eta\rightarrow 0$.
Using $\ket{ C^-_{\alpha_1,\alpha_2}} =  (\ket{\alpha_1,\alpha_2}-\ket{-\alpha_1,- \alpha_2})/N^-_{\alpha_1,\alpha_2} $ and expanding the coherent states in terms of Fock states leads to

\begin{align}
 \ket{ C^-_{\alpha_1,\alpha_2}} N^-_{\alpha_1,\alpha_2}= \left[\sum_{n,m=0}^\infty \frac{\alpha_1^n\alpha_2^m}{\sqrt{n!m!}}\ket{n,m} \right.  \left. - \frac{(-\alpha_1)^n(-\alpha_2)^m}{\sqrt{n!m!}}\ket{n,m} \right]e^{-(|\alpha_1|^2+|\alpha_2|^2)/2} \nonumber
\end{align}

For $|\alpha_1|, |\alpha_2|\rightarrow 0$  and to first order in $\alpha_1, \alpha_2 $  we get

\begin{align}
 \ket{ C^-_{\alpha_1,\alpha_2}} N^-_{\alpha_1,\alpha_2} \approx \left[\ket{0,0}+\alpha_1\ket{1,0}+\alpha_2\ket{0,1}  - \ket{0,0}+\alpha_1\ket{1,0}+\alpha_2\ket{0,1} \right] \nonumber
\end{align}

Therefore depending on the exact path to 0 we get
\begin{align}
 \ket{ C^-_{\alpha_1,\alpha_2}} \rightarrow \lim_{\eta \rightarrow 0} \frac{\alpha_1(\eta)\ket{1,0}+\alpha_2(\eta)\ket{0,1}}{\sqrt{|\alpha_1(\eta)|^2+|\alpha_2(\eta)|^2}}.
\end{align}

Similarly  we get
\begin{align}
 \ket{ C^-_{\alpha_1,-\alpha_2}} \rightarrow \lim_{\eta \rightarrow 0} \frac{\alpha_1(\eta)\ket{1,0}-\alpha_2(\eta)\ket{0,1}}{\sqrt{|\alpha_1(\eta)|^2+|\alpha_2(\eta)|^2}}.
\end{align}

\section{derivation of the Berry phase}
We now derive the Berry phase that is generated by variations of the complex drive $\epsilon = r e^{i\phi}$.
For any adiabatic change of $\epsilon$ with handle $s_i$ where $s_1= r$ and $s_2 = \phi$, a system which is initially in the eigenstate $\ket{C^\pm_{\alpha(\vec{s}(0))}}$ with energy $E(\vec{s}(0))$ will at all times $t$ be in the instantaneous eigenstate $\lambda(t) \ket{C^\pm_{\alpha(\vec{s}(t))}}$, where $\lambda(t)$ is a phase and is given by 
\begin{align}
\lambda(t) = e^{-\frac{i}{\hbar}\int_0^t dt^\prime E(\vec{s}(t^\prime))} \cdot e^{i \varphi_B(t)}.
\end{align}
The first part of this expression is the dynamical phase accumulated by the system due to oscillations with the eigenenery of the instantaneous eigenstate, while the second term is the geometric phase that the system accumulates which is independent of the duration of the variation and only depends of the path that is taken in parameter space.
For a curve C the accumulated geometrical phase is given by
\begin{align}
 \varphi_B^\pm = - \mathrm{Im} \int_C d \vec{s} \braket{C^\pm_{\alpha(\vec{s})}|\vec{\nabla}_{\vec{s}}C^\pm_{\alpha(\vec{s})}}.
\end{align}
Evaluating the integrand for general variations $s_i$, one obtains
\begin{align}
    \braket{C^\pm_{\alpha(\vec{s})}|\frac{\partial}{\partial_{s_i}}C^\pm_{\alpha(\vec{s})}} 
= \left( \frac{\partial \alpha}{\partial s_i} \alpha^\star- \frac{\partial \alpha ^\star}{\partial s_i} \alpha \right) \frac{1}{2} \left[  \frac{1\mp e^{-2|\alpha|^2}}{1\pm e^{-2|\alpha|^2}}\right].
\end{align}
Inserting $\alpha = \sqrt{r/K}e^{i\phi /2}$ we get $\left( \frac{\partial \alpha}{\partial r} \alpha^\star- \frac{\partial \alpha ^\star}{\partial r} \alpha \right) = 0 $ and $\left( \frac{\partial \alpha}{\partial \phi} \alpha^\star- \frac{\partial \alpha ^\star}{\partial \phi} \alpha \right) = i |\alpha|^2 $. This leads to

\begin{align}
    d \vec{s} \braket{C_{\alpha(\vec{s})}^{\pm}|\vec{\nabla}_{\vec{s}}C_{\alpha(\vec{s})}^{\pm}} 
= i |\alpha(\phi)|^2  \frac{1}{2} \left[  \frac{1\mp e^{-2|\alpha(\phi)|^2}}{1\pm e^{-2|\alpha(\phi)|^2}}\right] d\phi.
\label{equation_berry_phase}
\end{align}
The accumulated Berry phase therefore is independent of variations of the drive strength and only depends on variations of the phase of the two photon drive.

If one considers a variation of the drive where the drive strength is kept constant but the phase of the drive is slowly varied so that the drive is slowly rotating around the origin so that $d\vec{s} = d\phi$, and $|\alpha|$ is kept constant during the variation, the Berry phase accumulated during a full rotation of the phase of the drive is given by
\begin{align}
 \varphi_B^\pm =-|\alpha|^2  \frac{1}{2} \left[  \frac{1\mp e^{-2|\alpha|^2}}{1\pm e^{-2|\alpha|^2}}\right] 2 \pi .
 \label{eq:phi_B_1_rot}
\end{align}
As one can see the Berry phase accumulated in the even and odd subspace are different from each other while the dynamical phase acquired by the even and odd state are the same as both states are degenerate in energy. This stems from the fact that coherent states form an overcomplete basis, leading to a nonzero overlap between $\ket{\alpha}$ and $\ket{-\alpha}$. In fact the overlap between two coherent states $\ket{\alpha}$ and $\ket{-\alpha}$ is given by $\braket{\alpha|-\alpha} = e^{-2 |\alpha|^2}$, which is the term in equation \ref{eq:phi_B_1_rot} that distinguishes the Berry phases of the even and the odd subspace. Since the overlap between two coherent states decreases the farther they are apart in phase space, the overlap $\braket{\alpha|-\alpha}$ and therefore the difference of Berry phase for $\ket{C^\pm_\alpha}$ goes to 0 for $|\alpha| \gg 1$.

\begin{figure*}
	\begin{center}
	\includegraphics[width=0.8\linewidth]{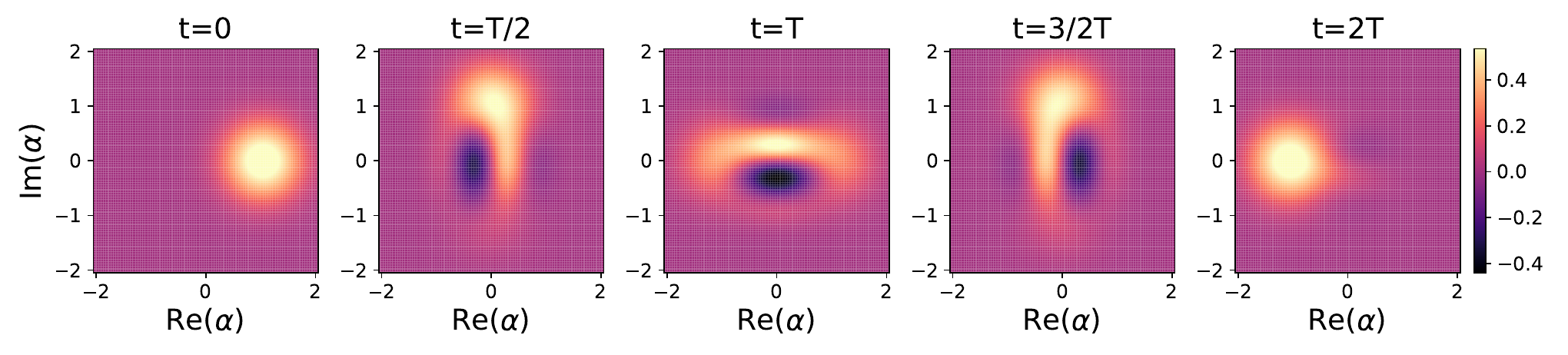}
       \includegraphics[width=0.8\linewidth]{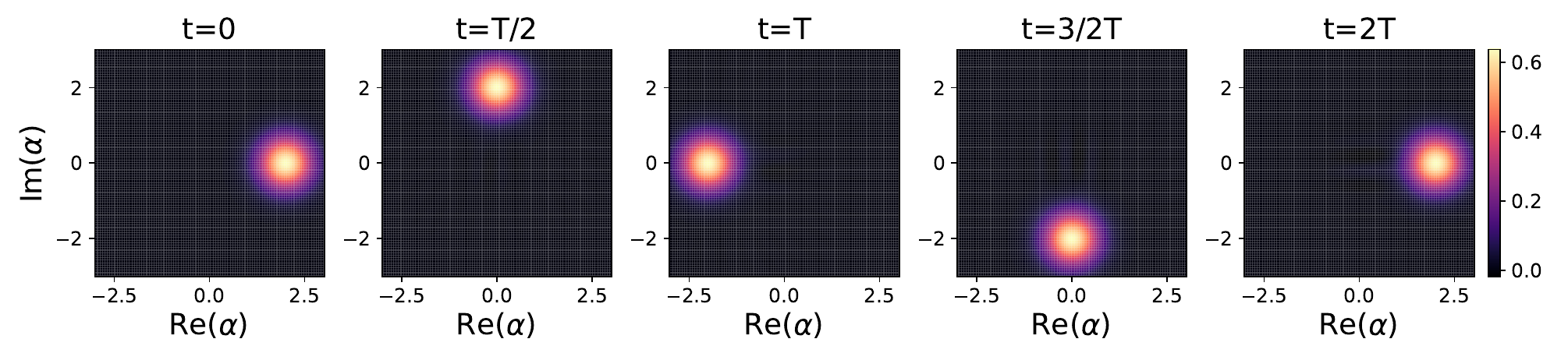}
	\end{center}
	\caption{Evolution of the Wigner function of the state of a system that was initialized in $\Psi(0)=\ket{\alpha}$. The simulation is done for $\alpha=1.04$ (top) and $\alpha=2$ (bottom). In both cases the fidelity of the final state is $>$ 0.99. The time T that the phase of the drive takes to do one full rotation is given by T=50 1/K.} 
\label{fig:transition_wigner}
\end{figure*}
\section{Wigner function of the rotation of a single mode KPO}
In figure \ref{fig:transition_wigner} (top) the Wigner function of the state of the resonator which was initially in $\ket{\alpha =1.04}$ is plotted at different times during the rotation of $\epsilon $. One can see that as expected after two rotations of $\epsilon$ the final state of the system is given by $\ket{\alpha=-1.04}$. 
As a comparison the Wigner function of a state that was initialize in $\ket{\alpha}=2$ is plotted in figure \ref{fig:transition_wigner} (bottom). One can see that for bigger $\alpha$ the difference in phase for a small number of rotations is negligible so that after two rotations of the drive the state is approximately given by $\ket{\alpha = 2}$ again. 
In both simulations the fidelity of the final state with respect to the analytical expectation is $>0.99$ indicating that the rotation of the drive was slow enough to be considered adiabatic.
Furthermore by looking at intermediate steps of the time evolution one can see that the state does not just perform a rotation in phase space, but actually evolves through a superposition of $\ket{\alpha}$ and $\ket{-\alpha}$.

\end{document}